\documentclass[aps,prb,twocolumn,superscriptaddress]{revtex4-2}
\usepackage{graphics}
\usepackage{longtable}
\usepackage{epsfig}
\usepackage{amsmath}
\usepackage{float}
\usepackage{placeins}
\usepackage{color,soul}
\begin{document}
\bibliographystyle{apsrev}
\title{The current magnetization hypothesis as a microscopic theory of the \O{}rsted magnetic field induction}
\author{Sherif Abdulkader Tawfik}
\email{sherif.tawfic@gmail.com}
\affiliation{Institute for Frontier Materials, Deakin University, Geelong, Victoria 3216, Australia.}
\affiliation{ARC Centre of Excellence in Exciton Science, School of Science, RMIT University, Melbourne, VIC 3001 Australia.}
\begin{abstract}

A wire that conducts an electric current will give rise to circular magnetic field (the \O{}rsted magnetic field), which can be calculated using the Maxwell-Ampere equation. For wires with diameters in the macroscopic scale, the Maxwell-Ampere equation is an established physical law that has can reproduce a range of experimental observations. A key implication of this equation is that the induction of \O{}rsted magnetic field is only a result of the displacement of charge. A possible microscopic origin of \O{}rsted magnetic induction was suggested in [J. Mag. Mag. Mat. 504, 166660 (2020)] (will be called the current magnetization hypothesis (CMH) thereupon). The present work establishes computationally, using simplified wire models, that the CMH reproduces the results of the Maxwell-Ampere equation for wires with a square cross section. I demonstrate that CMH could resolve the apparent contradiction between the observed induced magnetic field and that predicted by the Maxwell-Ampere equation in nanowires, as was reported in [Phys. Rev. B 99, 014436 (2019)]. The CMH shows that a possible reason for such contradiction is the presence of non-conductive surface layers in conductors.

\end{abstract}

\maketitle

\section{Introduction}

One of the most established laws in science is the Maxwell-Ampere equation. Since the introduction of the original Ampere's equation in 1820 to explain the magnetic field induction phenomenon discovered by \O{}rsted, and then its reformulation by Maxwell as the Maxwell-Ampere equation, the latter has been shaping our understanding of the nature of the relationship between the magnetic and electric fields. Coupling it with the Faraday equation, Maxwell obtained a beautifully symmetrical relationship between the electric and magnetic fields, a symmetry that was later expressed by Einstein in a compact form in terms of the electromagnetic tensor. The duet of these two equations permeates nearly all branches of physics.


However, the application of the Maxwell-Ampere equation, or its solution the Biot-Savart equation, for current-carrying structures with nano-scale and/or reduced dimensionalities is challenging for three reasons: (1) it is computationally difficult to generate the \O{}rsted magnetic field for structures that have irregular shape, (2) the application of the equation to structures with low dimensionalities, such as two-dimensional (2D) materials, suffers from fundamental inaccuracies, as I commented in Ref. \cite{Sherif} on the application of the 2D Maxwell equations to predict the \O{}rsted magnetic field of graphene in Ref. \cite{NV}, and (3) it is difficult to perform measurements of the \O{}rsted magnetic field at such length scales in order to validate the theoretical predictions. 

The first such measurement was reported by Tetienne \textit{et al.} \cite{NV_Problem}. However, they showed that the observed \O{}rsted magnetic field in nanostructures is significantly different from the field derived from the Maxwell-Ampere equation. While the validity of the Maxwell-Ampere equation for macroscopic wires with a circular cross-section is beyond question, and assuming the accuracy of the measurements performed in Ref. \cite{NV_Problem}, the observations reported in Ref. \cite{NV_Problem} can be explained either (i) by stretching the assumptions in the model to reconcile the measurements with the Maxwell-Ampere  equation as applied to a \textit{modified} model, as was done by the authors, or (ii) by deriving a microscopic theory for the \O{}rsted magnetic field induction which agrees with the form of the Maxwell-Ampere  equation at the macroscopic scale, but departs from it at the microscopic scale. Option (i) is the safer and more compliant of the two options, and the second option is challenging because it will involve the introduction of possibly new physics.

I have ventured to choose option (ii) in my recent work, Ref. \cite{Sherif}. In this paper I have made the following proposal: the \O{}rsted magnetic field emerges from each conduction electron, rather than emerging from the totality of the \textit{continuum} electric current density (it is worth noting here that the Maxwell-Ampere equation is still a \textit{continuum theory}). I shall call the proposed theory the current magnetization hypothesis (CMH). This is a more appropriate name than the name introduced in Ref. \cite{Sherif} (the generalized spin-orbit interaction (GSOI)), and is related to the name ``wire magnetization hypothesis'' which was introduced by Biot and Savart \cite{Assis}. A key advantage in this model is that it helps alleviate challenges (1) and (2) as follows: given that the \O{}rsted magnetic field is calculated from first principles based on the atomic structure of the conducting material, the field can be calculated for structures with arbitrary shapes, sizes and dimensionalities. Hence, the inconsistencies of the Maxwell equations at reduced dimensionalities will be avoided.

Note that there is already a microscopic version of the Biot-Savart equation for a moving electron, which is the well known relationship $\textbf{B}(\textbf{r}) = \frac{1}{c^2} \textbf{v} \times \textbf{E}(\textbf{r})$. This equation establishes that (1) the electron alone induces the \O{}rsted magnetic field when it moves with a non-zero speed, and that (2) the \O{}rsted magnetic field decays as $1/r^2$. The generation of the Biot-Savart form of the magnetic field induced by a current carrying wire, which decays as $1/r$, is derived from the integration of the Biot-Savart field of the electron. In CMH, as I suggested in Ref. \cite{Sherif}, an alternative induction mechanism is suggested: each individual electron produces its own \O{}rsted magnetic field which decays as $1/r$, as long as it moves relative to other charged particles. The CMH field of a current-carrying wire is therefore a direct aggregation of these fields.

In this work, I present the CMH as a microscopic theory for the \O{}rsted magnetic field induction, and demonstrate its predictions in comparison to those of the Biot-Savart equation for wires with square cross sections. Then, I examine the application of the CMH for wires with different dimensionalities, including a 2D sheet and an atomic wire.

\section{The current magnetization hypothesis}

According to CMH, the \O{}rsted magnetic field is induced as a result of a two-body interaction: two charged particles moving relative to each other at a non-zero relative velocity. Inside a conductor, the flow of current is essentially a movement of electrons relative to the static metal ions. A CMH magnetic field is induced at the position of each moving electron, and the totality of the CMH magnetic fields of those electrons constitute the \O{}rsted magnetic field in conductors. Here I present an application of the CMH for a range of nanostructures by numerical computations of the field using simple dynamical models, where electrons are treated classically. I show that, while the CMH agrees with the Maxwell-Ampere equation in predicting the \O{}rsted magnetic field in conductors with nanowires with a square cross-section, the CMH predicts smaller \O{}rsted magnetic field in nanowires with rectangular cross-sections. The CMH prediction for \O{}rsted magnetic field of a single atom chain (SAC) is $\sim 1/30$ of the prediction of the Biot-Savart equation. 

In the CMH, the magnetic field induced by the motion of a charged particle relative to another charged particle, $\textbf{B}_{GSOI,ij}$, is given by

\begin{eqnarray}
\label{eq:GSOI}
\textbf{B}_{SOI,i}&=&-\frac{1}{c^2}(\textbf{v}_i-\textbf{v}_j)\times \textbf{E} \quad,\\ \nonumber
&=&-\frac{1}{c^2}(\textbf{v}_i-\textbf{v}_j)\times q_j\frac{\textbf{r}_i-\textbf{r}_j}{\left| \textbf{r}_i-\textbf{r}_j\right| ^3} \quad,\\
\textbf{B}_{GSOI,ij}(\textbf{r})&=&  \textbf{B}_{SOI,ij}  B_{d,ij}B_{r}(\left| \textbf{r}_i-\textbf{r}\right|) \quad, \\
B_{d,ij}&=&e^{-k\left| \textbf{r}_i-\textbf{r}_j\right|} \quad,  \\
B_{\rm{rad}}(r) &=& \frac{ \kappa }{ r +\kappa}
\end{eqnarray}

\noindent where $q_j$ is the charge of particle $j$, $B_{d,ij}$ is the exponential decay of the electric field inside a conductor due to charge screening, and $B_{\rm{rad}}(r)$ represents the radiation decay of the point magnetic field arising due to SOI. In this equation, $\textbf{B}_{SOI,ij}$ is proportional to $q_j$, which is the effective nuclear charge of the ion $j$ in the conductor. $\kappa$ in Eq. \ref{eq:GSOI} is an empirical quantity, which determines the decay of the SOI magnetic field. The value that has been obtained in this work by fitting is 5$\times 10^{-3}$~\AA. The introduction of the decay factor, $B_{\rm{rad}}(r)$, was not derived from more fundamental theories, but was based on substituting the observed decay of the \O{}rsted magnetic field of a current-carrying wire ($1/r$ according to the Biot-Savart equation) into a microscopic theory.

With regards to the decay factor $B_{\rm{rad}}(r)$, this term becomes 1 when we calculate the value of the $B_{\rm{rad}}$ field induced by particle $i$ at the position of particle $i$ (that is, itself). This is unlike Coulomb's law for the electric charge or the magnetic monopole, in which the field magnitude goes to infinity as $r$ approaches the electric/magnetic charge source.

\section{Nanowires with a square cross-section}

\begin{figure}[h]
	\includegraphics[width=90mm]{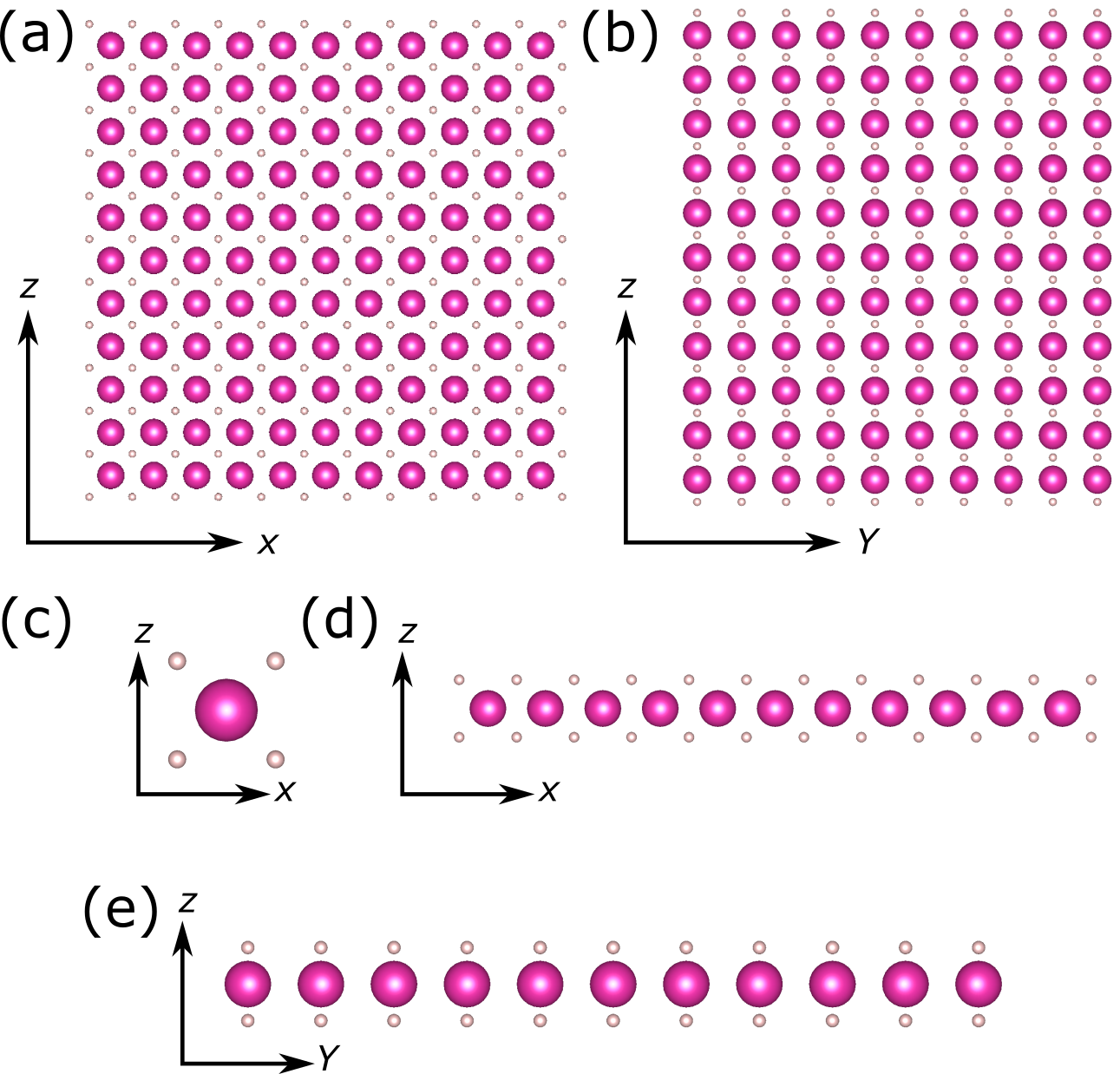}
	\caption{\label{fig2} The structure of the conducting nanostructures, where pink spheres represent atoms and white spheres represent conduction electrons. (a) Front and (b) side views of a  nanowire with a square cross-section. (c) Front views of (c) an atomic wire and (d) a thin nanoribbon, such as graphene. (e) Side view for the structures in (c) and (d).}
\end{figure}

First, I demonstrate the application of Eq. \ref{eq:GSOI} to one-dimensional nanostructures with rectangular cross section. The predictions of the CMH for square conductors are compared against the predictions of the Biot-Savart equation to show that they are qualitatively in agreement. The conductor is assumed to have a simplistic atomic structure, with a simple cubical symmetry, as shown in Figure \ref{fig2}(a). Each unit cell accommodates a single conducting electron, and each atom has an effective nuclear charge of 5.842 (that of copper). This approximate approach was applied earlier for modeling the Ampere tension in conducting wires \cite{Tension3}, and approximates the symmetry of transition metal conductors such as Cu, Au and Ag, where the outermost $s$ electron is responsible for its metallic conductivity. The lattice constant of the unitcell is 3~\AA, and the electrons are assumed to flow in the midpoint between each two neighboring atoms along the $y$-axis. The nanowires are 10,000 atoms long (3 $\mu$m). The Biot-Savart  \O{}rsted magnetic field is obtained by the analytical expressions in Ref. \cite{BS_Rectangular}. 

\begin{figure}[h]
	\includegraphics[width=90mm]{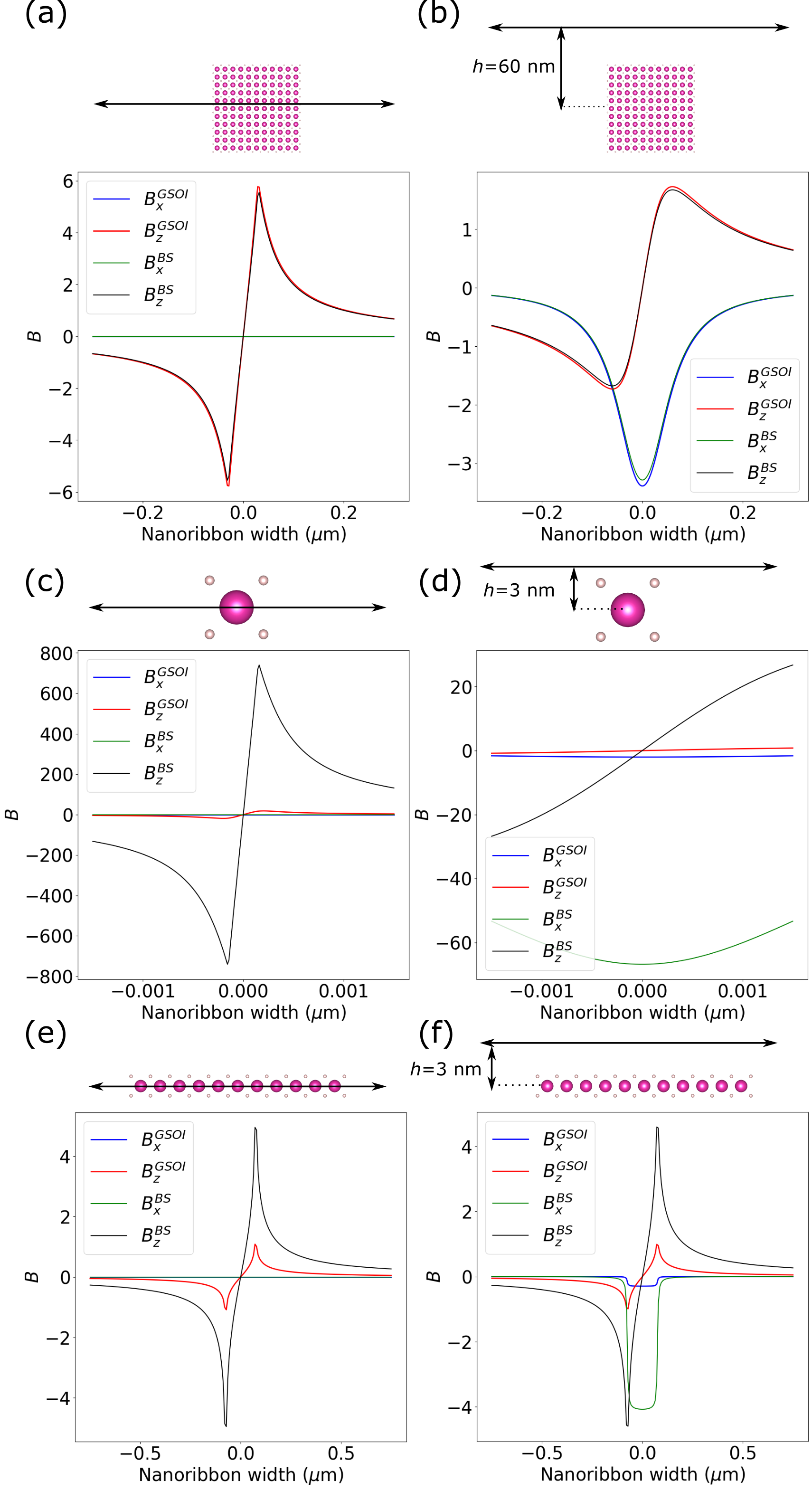}
	\caption{\label{fig_Comparisons} Comparison between the \O{}rsted magnetic field calculated using Biot-Savart (BS) and CMH (using the GSOI equation, Eq. \ref{eq:GSOI}) for a nanowire with (a,b) a square cross-section with $201\times 201$ atoms, (c,d) a single atom chain, or atomic wire, and (e,f) an atomically thin sheet with $501\times 1$ atoms. The $B^{BS}_x$ and $B^{BS}_z$ curves are the  components of $\textbf{B}_{BS}$ using the formulas in \cite{BS_Rectangular} for wires with rectangular cross-sections, and the of $B^{GSOI}_x$ and $B^{GSOI}_z$  curves are the  components of $\textbf{B}_{GSOI,ij}$, multiplied by 1.3. In (a,c,e), the \O{}rsted magnetic field is calculated along the axis indicated above the figure, which passes through the center of the wire. In (b,d,f), the axis lies $h$ above the center.}
\end{figure}

For the calculation of the \O{}rsted magnetic field using the CMH, the relationship $I = v\rho A$ is used to find the value of $v$, the speed of the conduction electron, given that the following values are given: $I$, the electric current, $\rho$, the electron density of the conductor (C/m$^3$), and $A$, the area of the conductor's cross section. To compare CMH with the Biot-Savart equation, the values of $I$ and $\rho$ will be held fixed, while $A$ will change as the conductor size changes. Thus, the value of $v$ that will be substituted into Eq. \ref{eq:GSOI} will be 
\begin{equation}
v=I/\rho A \quad.
\label{eq:v}
\end{equation} 
\noindent The total \O{}rsted magnetic field at every point along the horizontal axis is the sum of the $\textbf{B}_{GSOI,ij}$ from all of the electrons in the wire. The nanowire in Fig. \ref{fig_Comparisons}(a,b) has $201 \times 201 \times 10000$ atoms (and therefore $202 \times 202 \times 10000$ electrons), which is a huge calculation. I have written a python version 3.9 code that parallelizes this calculation and therefore speeds it up. The the \O{}rsted magnetic field calculation of a structure, in which the cross-sectional area (in terms of number of electrons) is $X\times Z$ and is $Y$ long,  progresses over two steps: first, a lookup table of $\textbf{B}_{SOI,i}$ values for each of the $X\times Z$ electrons is calculated; second,  $\textbf{B}_{GSOI,ij}$ is calculated by summing the contribution of every electron in each of the cross-sectional slices along the length of the wire.

Fig. \ref{fig_Comparisons}(a,b) shows that, except for a factor of $\sim 1.3$, the \O{}rsted magnetic field using the CMH is very close to that of Biot-Savart for various wire cross-sectional sizes. Given that this 30\% difference between the CMH and Biot-Savart is consistent for the different cross-section sizes examined here ($51 \times 51$, $101 \times 101$, $151 \times 151$ and $201 \times 201$ atoms), it will be applied throughout this work for predicting the \O{}rsted magnetic field of other structures. This result validates the CMH as an alternative theory for the derivation of the  \O{}rsted magnetic field of conductors.

\begin{figure}[h]
	\includegraphics[width=70mm]{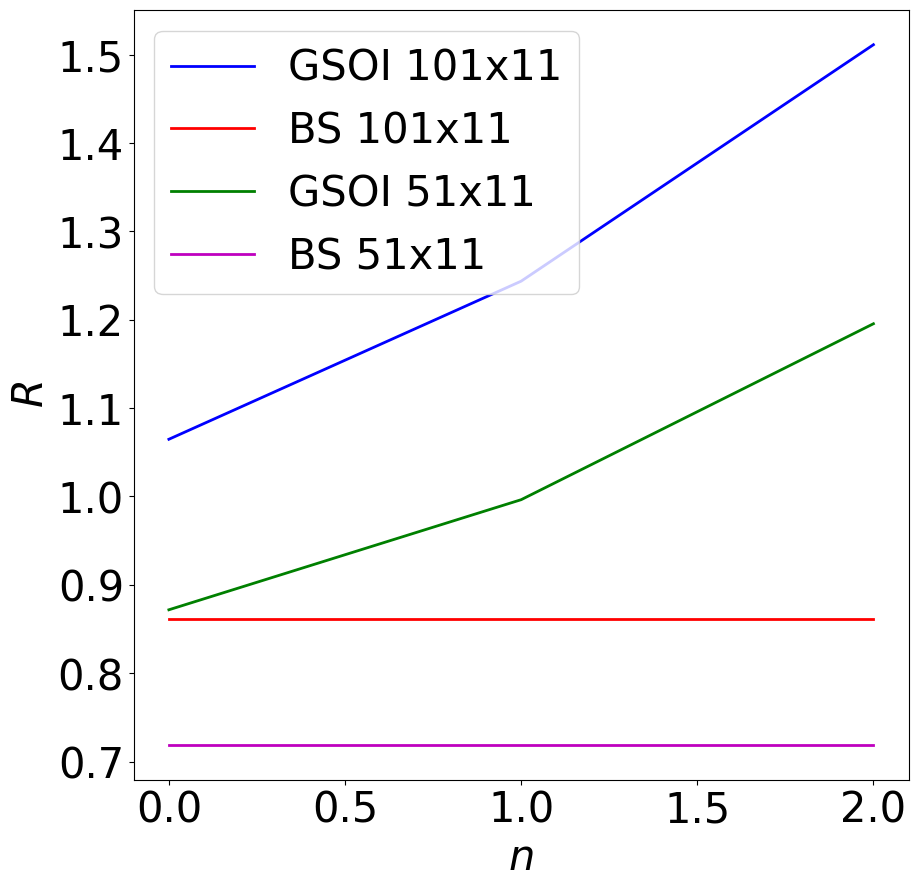}
	\caption{\label{R_n} Comparison between the $R_n$ ratio calculated for the $51 \times 11$ and $101 \times 11$ nanowires using CMH (calculated using the GSOI equation, Eq. \ref{eq:GSOI}) and the Biot-Savart (BS) equation, where $h=3$ nm.}
\end{figure}

Moreover, CMH can qualitatively explain the reduction of the observed $B_x$ component of the \O{}rsted magnetic field in thin nanowires compared with $B^{BS}_x$ \cite{NV_Problem} as arising from the presence of non-conductive surface layers. These surface layers occur in metallic conductors, such as the patinated surface layer (semiconducting copper oxide layer) in copper wires. For a rectangular conductor with a cross section of $101 \times 11$,  and $h=3$ nm (where $h$ is defined in Fig. \ref{fig_Comparisons}), the ratio $R={\rm max}(\left| B_z \right|)/{\rm max}(\left| B_x \right|)$ is calculated. Here ${\rm max}(\left| B \right|)$ denotes the maximum value of $\left| B \right|$. The value of $R$ for this structure is $\sim 1.1$ using the CMH and $\sim 0.9$ using the Biot-Savart equation. I then model presence of $n$ non-conductive layers around the nanowire by removing the  outermost $n$ layer of electrons surrounding the structure. The values of $R$ are plotted against $n$ (denoted $R_n$) in Fig. \ref{R_n}. The experimental measurement reported in Ref. \cite{NV_Problem} shows that $B_x$ is $\sim 94\%$ less than $B^{BS}_x$. The trend exhibited in Fig. \ref{R_n} indicates that, as the nanowire width increases, $R_0$ will increase, and as $n$ increases, $R_n$ will increase. The depletion of $B_x$ can at east be qualitatively implied, in line with the observation in Ref. \cite{NV_Problem} .

\section{Single atom chains and atomically thin sheets}

Next, the CMH is applied to low-dimensional nanostructures, including SACs and atomically thin sheets. SACs have been realized using various transition metal atoms, and have been a notable platform for conductance quantization \cite{AuSAC,AuSAC2}. An SAC is assumed here to have the structure in Fig. \ref{fig2}(c): a column of atoms, where each atom is surrounded by 4 electrons. Thus, for the purpose of calculating the \O{}rsted magnetic field using the Biot-Savart equation, this system is not modeled as in infinitely thin conductor. Instead, it is modeled as a very thin conductor with a square cross-section that has a side length of 3~\AA. Therefore, the electron velocity is calculated using Eq. \ref{eq:v} instead of the linear current density relationship $v=I/l$, where $l$ is the linear charge density (C/m).

Fig. \ref{fig_Comparisons}(c,d) shows the comparison between the CMH and Maxwell-Ampere predictions for the atomic wires. The \O{}rsted magnetic field predicted by the CMH is $\sim 1/30$ of that of the Biot-Savart equation, and such large difference between the two theories would naturally require an experimental benchmark, which would support one of the two theories, or neither of them. However, \textit{measuring the \O{}rsted magnetic field induced by a conducting atomic wire with such lengths has never been reported}. This is possibly because a wire that is long enough to allow for the measurement of the induced magnetic field has not yet been fabricated.


For the case of atomically thin sheets, an example of such structures is graphene. In this material, conductance mainly takes place through the two $\pi$ layers of the hexagonal network, whereas the $\sigma$ electrons are mostly localized and do not participate in conductance. Thus, a graphene sheet can be modeled as a bilayer conductor, as displayed in Fig. \ref{fig2}. The structure of the electric current in graphene is therefore fundamentally different from an infinitely thin conductor, where the current is constrained in 2D. Therefore, using the 2D limit of the Biot-Savart equations to model the \O{}rsted magnetic field of graphene is not valid. Moreover, as pointed out in Ref. \cite{Sherif}, such 2D limit is numerically unreliable. Therefore in the comparison between the CMH and Biot-Savart for atomically thin sheets, a 3~\AA-thin sheet is used for calculating the \O{}rsted magnetic field using the Biot-Savart equation.

Fig. \ref{fig_Comparisons}(e,f) show that the predictions of the CMH for both $x$ and $z$ components of the \O{}rsted magnetic field are smaller than those of the Biot-Savart equation: in Fig. \ref{fig_Comparisons}(e), $B^{GSOI}_z$ is $\sim 1/5$ of $B^{BS}_z$, and in \ref{fig_Comparisons}(f), $B^{GSOI}_x$ is $\sim 1/5$ of $B^{BS}_x$ and $B^{GSOI}_x$ is $\sim 1/15$ of $B^{BS}_x$. Interestingly, in Fig. \ref{fig_Comparisons}(f), the $R$ ratio using CMH is $\sim 3.4$, and for Biot-Savart is $\sim 1.1$. Such large difference between the CMH and Biot-Savart predictions, in general, highlights that the prediction of an \O{}rsted magnetic field as high as 1 T in graphene nanosolenoids, as reported in Ref. \cite{NEM1}, might be an overestimation.

\section{Conclusion}

This work sheds light on the knowledge gap in magnetic nanoinductors: the gap keeps widening as the nanoinductor becomes smaller because of the difficulty of observing the \O{}rsted magnetic field at such length scales. As soon as magnetic induction measurements started to be conducted for nano-systems, the measurement results either contradicted established theory (the Maxwell-Ampere equation), as in Ref. \cite{NV_Problem}, or the mathematical framework that was applied for explaining the observations was not free from problems, as I commented in Ref. \cite{Sherif} on the application of the 2D Maxwell equations to predict the \O{}rsted magnetic field of graphene in Ref. \cite{NV}. There is a rising interest in nano-electromagnetic systems: the miniaturization of the inductor circuit component into NEM structures is essential for the incorporation of inductors into integrated circuits, and graphene-based inductors have recently been theoretically investigated \cite{NEM1} and experimentally fabricated \cite{NEM2,NEM5}. NEMs have also been demonstrated as nano-robotic arms for moving nano-sized objects \cite{NEM3,NEM6} and for controlling the spin state of quantum dots using the induced \O{}rsted magnetic field \cite{NEM4}. Thus, critical examination of established magnetic induction theories is becoming increasingly urgent.

In summary, this work demonstrates the application of the current magnetization hypothesis as a microscopic origin for the \O{}rsted magnetic field induction in electrically conducting nanostructures. Using numerical calculations on simplified models for nanowires, it is shown that \O{}rsted magnetic field obtained using the current magnetization hypothesis converges to that obtained using the Maxwell-Ampere equation. The calculations show that the \O{}rsted magnetic field induced in nanomaterials, including single atom wires and atomically thin sheets, is much smaller than that predicted using the Biot-Savart equation. Importantly, for thin nanowires with rectangular cross section, the current magnetization hypothesis prediction of the planar component of the \O{}rsted magnetic field is in qualitative agreement with experimental measurements. Thus the current magnetization hypothesis is suggested as a microscopic theory of magnetic induction.

\section{Data Availability}

The data that support the findings of this study are available from the author upon reasonable request.

\end{document}